\documentclass[conference]{IEEEtran}
\usepackage{mathtools, cuted} 
\usepackage{lettrine} 
\usepackage{amsthm,amsmath} 
\usepackage{epsfig,amssymb,amsbsy,verbatim,subfigure,array}
\usepackage{pstricks,cite,url}
\usepackage{multicol,afterpage,wrapfig,paralist} 
\usepackage{hyperref}
\usepackage{breakurl}
\usepackage{verbatim,tikz,subfigure}
\usepackage{tikz,pgfplots} 
\usepackage[english]{babel}
 
{}

\def\d{\mathrm{d}}

\title{Coverage Analysis of a Thinned LiFi Optical Attocell Network}
\author{\IEEEauthorblockN{Atchutananda Surampudi}
\IEEEauthorblockA{Department of Engineering Science, University of Oxford, OX1-3PJ, Oxford. \\
atchutananda.surampudi@wadham.ox.ac.uk}
}
\begin{document}
\maketitle

\begin{abstract}
This work analyzes coverage in the downlink of a thinned LiFi attocell network of deterministically arranged LEDs.     The network is thinned by a Bernoulli probability $p$ over all the LEDs to decide whether each one of them acts as a LiFi source or not. Then we use the series approximation approach used in \cite{atc1} to obtain closed form expressions for the probability of coverage in such thinned LiFi attocell networks and validate them using numerical simulations. 
\end{abstract}

\section{Introduction}
Light Fidelity (LiFi) has emerged as a high speed wireless data access solution using visible light \cite{hass}. 
For downlink access, the arrangement of a network of LiFi sources using light emitting diodes (LEDs) is called an attocell network. Such an attocell network is centrally monitored and is generally arranged in a deterministic lattice. 
In many situations, for example in a large conference room or a library, out of such a deterministic lattice arrangement, all the LEDs providing illumination may not be acting as LiFi data access points. Those LEDs that act as LiFi sources are then randomly located over a thinned version of the original deterministic lattice. Modelling such a random point process of LiFi-LEDs has been an open area of research. The Poisson point process (PPP) that is generally used to model conventional wireless networks, cannot be assumed in this case because it does not appropriately consider into account the minimum separation between the randomly located points over a deterministic lattice. 
The corresponding analysis of the signal-to-interference-plus-noise-ratio (SINR) and the determination of the probability of coverage at a particular receiver also becomes difficult since the time varying fading over the line-of-sight LiFi channels is considered to be absent. 

 
\subsection{Related works}
This problem of modelling such a point process of LiFi-LEDs and the analysis of the corresponding probability of coverage has been cited in \cite{cov1, lit1, lit2, lit3}. Almost all of them assume a PPP of LiFi-LEDs and derive tractable expressions to analyze coverage under no-fading conditions. But as mentioned earlier, this assumption does not appropriately take into account the minimum separation between the LEDs over the deterministic lattice. 
Closed form expressions for co-channel interference and SINR have been derived in \cite{atc1}, but only for the  case when all the LEDs are LiFi sources over the lattice without randomness.
\subsection{Our contributions} 
This work proposes a novel solution to characterize coverage in a thinned LiFi attocell network. Firstly, every LED in the network is assigned a Bernoulli probability $p$ of acting as a LiFi source. Then the interference is modelled as a an infinite summation over a set of weighted Bernoulli random variables that are independent. This large sum is approximated to be converging in distribution to a Gaussian random variable whose mean and variance are exactly calculated. The series approximation approach used in \cite{atc1} is implemented to provide closed form expressions for the mean and the variance, which eventually are used to calculate an exact expression for the probability of coverage over an attocell.  

\subsection{Arrangement of the paper}
The paper is arranged as follows. Section II describes the system model. Section III is the main technical part of the paper that derives the expression for the probability of coverage. Validations using numerical simulations are presented in Section IV. The paper concludes with Section V.         

\section{System model}
\subsection{The attocell network}
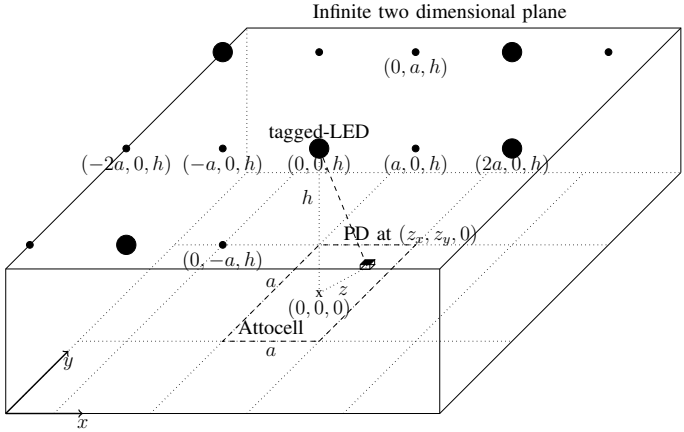
\begin{figure}[ht]  
\centering
 \resizebox{0.5\textwidth}{!}{%
 \begin{tikzpicture}
 \draw (0,0) -- (0,3) -- (5,8) -- (14,8) -- (14,5);
 \draw (14,5) -- (9,0) -- (0,0);
 \draw (14,8) -- (9,3) -- (0,3);
 \draw (9,0) -- (9,3);
 \draw [dotted] (0,0) -- (5,5) -- (14,5);
 \draw [dotted] (5,5) -- (5,8);
 
 \draw [line width= 0.03cm,->] (0,0) -- (1.6,0);
 \node [below] at (1.6,0) {\large $x$};
 \draw [line width= 0.03cm,->] (0,0) -- (1.3,1.3);
  \node [below] at (1.3,1.3) {\large $y$};
 
 \draw [fill] (0.5,3.5) circle [radius=0.07];
 
 \draw [fill] (2.5,3.5) circle [radius=0.2];
 
 \draw [fill] (4.5,3.5) circle [radius=0.07];
 \node [below] at (4.5,3.5) {\large $(0,-a,h)$};
 
 
 
 \draw [fill] (2.5,5.5) circle [radius=0.07];
 \node [below] at (2.5,5.5) {\large $(-2a,0,h)$};
 \node at (5.5,1.8) {\large Attocell};
 
 \draw [fill] (4.5,5.5) circle [radius=0.07];
 \node [below] at (4.5,5.5) {\large $(-a,0,h)$};
 
 \draw [fill] (6.5,5.5) circle [radius=0.20];
 \node [below] at (6.5,5.5) {\large $(0,0,h)$};
 \node [above] at (6.5,5.5) {\large tagged-LED};
 
 \draw [fill] (8.5,5.5) circle [radius=0.07];
 \node [below] at (8.5,5.5) {\large $(a,0,h)$};
 
 \draw [fill] (10.5,5.5) circle [radius=0.2];
 \node [below] at (10.5,5.5) {\large $(2a,0,h)$};
 
 \draw [fill] (4.5,7.5) circle [radius=0.2];
 
 \draw [fill] (6.5,7.5) circle [radius=0.07];
 
 \draw [fill] (8.5,7.5) circle [radius=0.07];
 \node [below] at (8.5,7.5) {\large $(0,a,h)$};
 
 \draw [fill] (10.5,7.5) circle [radius=0.2];
 
 \draw [fill] (12.5,7.5) circle [radius=0.07]; 
 
 \draw [dotted] (1,0) -- (6,5);
 \draw [dotted] (3,0) -- (8,5);
 \draw [dotted] (5,0) -- (10,5);
 \draw [dotted] (7,0) -- (12,5);
 \draw [dotted] (1.5,1.5) -- (10.5,1.5);
 \draw [dotted] (3.5,3.5) -- (12.5,3.5);
 \draw [dotted] (6.5,2.5) -- (6.5,5.5);
 \node [align=center] at (6.5,2.5) {\small x};
 \node [below] at (6.5,2.5) {\large $(0,0,0)$};
 \node [left] at (6.5,4.5) {\large $h$}; 
 \draw (7.35,2.99) -- (7.55,2.99) -- (7.65,3.09) -- (7.45,3.09) -- (7.35,2.99);
 \draw [fill] (7.35,3.09) -- (7.55,3.09) -- (7.65,3.19) -- (7.45,3.19) -- (7.35,3.09);
 \draw (7.35,2.99) -- (7.35,3.09);
 \draw (7.55,2.99) -- (7.55,3.09);
 \draw (7.65,3.09) -- (7.65,3.19);
 \draw (7.45,3.09) -- (7.45,3.19);
 \draw [dotted] (6.5,2.5) -- (7.5,3.04);
\node [below] at (8.4,4.04) {\large PD at $(z_{x},z_{y},0)$};
 \node [below] at (7,2.77) {\large $z$};
 \draw [dashed] (6.5,5.5) -- (7.5,3.04); 
 \draw [dashed] (6.5,1.5) -- (8.5,3.5) -- (6.5,3.5) -- (4.5,1.5) -- (6.5,1.5);
 \node [above] at (5.5,2.5) {\large $a$};
 \node [below] at (5.5,1.5) {\large $a$};
 \node [above] at (9,8) {\large Infinite two dimensional plane}; 
 
 \end{tikzpicture}
}%
\caption{This figure, adapted from \cite{atc1}, shows the infinite two dimensional LED network. There are infinite number of LEDs (circular dots) arranged symmetrically at regular intervals of $a$ all over the plane as a uniform square grid and installed at a height $h$. The rectangular dotted regions on ground depict the attocells corresponding to each LED above. Those LEDs which provide LiFi data access are depicted as relatively larger circles, while those which are idle are depicted as smaller ones. The user PD (small cuboid) at $(z_{x},z_{y},0)$ is assumed to have LiFi access from the tagged-LED at $(0,0,h)$ and the corresponding attocell is highlighted as dash-dot.}
\label{two_dim}
\end{figure}

Consider the infinite\footnote{An infinite network is considered in this work so as to model an ideal environment where the receiver receives interference from all directions and is located in the centre of the network. This assumption is practically equivalent to a large open room.} two dimensional arrangement of LEDs in Fig.\ref{two_dim}, all of them fixed at a height $h$, separated symmetrically by a distance $a$, and emitting light at a uniform average optical power $P_{o}$. Also, the LEDs have a Lambertian emission order $m = -\frac{\ln(2)}{\ln(\cos(\theta_{h}))}$, where $\theta_{h}$ is the half-power-semi-angle (HPSA) of any given LED. The photodiode (PD) of an area of cross section $A_{pd}$, responsivity $R_{pd}$ and field-of-view of $\frac{\pi}{2}$ radians is assumed to have its surface always parallel to the ground, i.e., without any orientation towards any LED, and is assumed to be located at $z=\sqrt{z_{x}^{2}+z_{y}^{2}}$ from the origin $(0,0,0)$. Importantly, the nearest LED at $(0,0,h)$ is assumed to emit data and the PD is tagged to this LED for LiFi access. For every $i^{th}$ LED, a random variable $\alpha_{i}$ is assigned that decides whether the LED acts as a LiFi source with probability $p$, or not with probability $1-p$. For brevity, $\alpha_{i}$ is clearly defined as follows.
\begin{align*}
\alpha_i \stackrel{d}{\sim} \left\{
               \begin{array}{ll}
                 p; \ \ \ \ \ \alpha_i=1,  \\
                 1-p; \alpha_i=0. \\ 
              \end{array}
              \right.
\end{align*}
This work neglects any non-linearities of the LED while intensity modulation. For the mathematical analysis, let all the LEDs belong to an infinite set $\mathbb{S}$. Let the modulation bandwidth of the system be $W$ and the noise power spectral density at the PD be $N_{o}$. Also, since the LEDs are assumed to be installed in a large open room, the multi-path and non-line-of-sight components received at the PD are considered relatively insignificant in power \cite{nlos1,nlos2,nlos3}.

\subsection{The signal-to-interference-plus-noise ratio}
Since all the LEDs saving the tagged LED are probably not LiFi enabled sources, the PD receives both unmodulated and modulated interference from the network. By assuming the value of $\alpha_{i}=0$ for the unmodulated sources, this work evidently oversees the unmodulated power received at the PD by assuming the fact that this unmodulated interference can be overcome by applying a suitable bias at the receiver or an equivalent high pass filter. Regardless of whether LiFi enabled or not, the optical intensities $s_{i}(t)$ from every $i^{th}$ LED experience a channel gain $G_{i}(z)$ given as \cite{atc1}
\begin{align*}  
G_{i}(z) =K(D_{i}^{2}+h^{2} )^{\frac{-(m+3)}{2}},
\label{eqn:gain2}   
\end{align*}
where $D_{i}$ represents the horizontal distance between the PD and the $i^{th}$ LED, and $K= \frac{(m+1)A_{pd}h^{m+1}}{2\pi}$. If $i=0$ represents the tagged LiFi enabled LED at $(0,0,h)$, then $D_{0}=z=\sqrt{z_{x}^{2}+z_{y}^{2}}$. All other LEDs $(i \in\mathbb{S}\setminus 0)$ now become co-channel interferers. The total received current $I(z,t)$ at the PD located at $z$ and during the time slot $t$ is given as 
\[ I(z,t) =s_{0}(t) G_{0}(z) R_{pd} + \mathcal{I}_{\infty}(z,t) + n(t),\]
where $n(t)$ is the additive white Gaussian noise process with power spectral density $N_{o}$, and a variance of $\sigma^{2}=N_{o}W$. Now, we can express the interference $\mathcal{I}_{\infty}(z,t)$ as
\[\mathcal{I}_{\infty}(z,t)=\sum_{i \in \mathbb{S}\setminus 0}\alpha_{i}s_{i}(t)G_{i}(z)R_{pd}.\]

So, after performing the time average over the received current, the electrical signal-to-interference-plus-noise ratio $\gamma(z)$ at the PD can be expressed as 
\begin{align}
\gamma(z)=\frac{P^{2}_{o}G^{2}_{0}(z)R^{2}_{pd}}{\sum_{i \in \mathbb{S}\setminus 0}\alpha_{i}P^{2}_{o}G^{2}_{i}(z)R^{2}_{pd}+\sigma^{2}}.
\label{eqn:sinr1}
\end{align}
The main results for the probability of coverage are presented in the following section.
\section{The probability of coverage} 
The probability of coverage $P_c(z,\theta)$, against a threshold $\theta$, can be expressed as a spatial average over the attocell dimensions as
\[ P_{c}=\frac{1}{a^{2}}\int_{-\frac{a}{2}}^{\frac{a}{2}}\int_{-\frac{a}{2}}^{\frac{a}{2}}P_{c}(z,\theta)\d z_{x}\d z_{y},  \]
where,
\begin{align}
P_{c}(z,\theta)&=\mathbb{P}[\gamma(z)>\theta|z], \nonumber\\
&\stackrel{(a)}{=}\mathbb{P}\bigg[\sum_{i \in \mathbb{S}\setminus 0}\alpha_{i}P^{2}_{0}G^{2}_{i}(z)R^{2}_{pd}<\frac{P^{2}_{o}G^{2}_{0}(z)R^{2}_{pd}}{\theta}-\sigma^{2}\bigg|z\bigg],\nonumber\\
&=\mathbb{P}\bigg[\sum_{i \in \mathbb{S}\setminus 0}\alpha_{i}(D_{i}^{2}+h^{2} )^{-\beta}<\eta\bigg|z \bigg], \nonumber\\
&=\mathbb{P}[C<\eta |z], 
\label{eqn:pc1}
\end{align}
where $(a)$ follows from rearranging \eqref{eqn:sinr1}; $\beta=m+3$, and $\eta=\frac{G^{2}_{0}(z)}{K^2\theta}-\frac{\sigma^{2}}{K^2P^{2}_{o}R^{2}_{pd}}$; and $C=\sum_{i \in \mathbb{S}\setminus 0}\alpha_{i}(D_{i}^{2}+h^{2} )^{-\beta}$ represents the infinite summation over a set of weighted $(w_{i}=(D_{i}^{2}+h^{2} )^{-\beta})$ independent and identical (iid) random variables, each of which is distributed as 
\begin{align*}
\alpha_{i}w_{i} \stackrel{d}{\sim} \left\{
               \begin{array}{ll}
                 p \ \ \ \ \  ;\alpha_{i}w_{i}=(D_{i}^{2}+h^{2})^{-\beta},  \\
                1-p; \alpha_{i}w_{i}=0, \\ 
              \end{array}
              \right.
\end{align*}
with the mean $\mu_{i}=\mathbb{E}[\alpha_{i}w_{i}]=p(D_{i}^{2}+h^{2})^{-\beta}$, and variance $\sigma_{1i}^{2}=\operatorname{Var}[\alpha_{i}w_{i}]=p(1-p)(D_{i}^{2}+h^{2})^{-2\beta}$. Correspondingly, the mean of $C$ can be expressed as 
\[ \mu=\mathbb{E}[C]=\sum_{i\in\mathbb{S}\setminus 0} p(D_{i}^{2}+h^{2})^{-\beta}, \]  
and the variance of $C$ can be expressed as
\[ \sigma_{1}^{2}= \operatorname{var}[C]=\sum_{i\in\mathbb{S}\setminus 0} p(1-p)(D_{i}^{2}+h^{2})^{-2\beta}. \]
Now, since the summation is over a large number of weighted iid random variables, $C$ is approximated to converge in distribution to a Gaussian with the mean $\mu$ and variance $\sigma_{1}^{2}$. This approximation is numerically validated with analytical simulations in section IV. As a matter of fact, $D_{i}^{2}=(ua+z_{x})^{2}+(va+z_{y})^{2}$, represents the horizontal distance of the PD from every other $i^{th}$ LED located at $(ua,va,h)$. So, the mean and variance can be expanded as 
$\mu=pS_{m}$, and $\sigma_{1}^{2}=p(1-p)S_{v}$, with 
\[ S_{m}=\sum_{u = -\infty}^{+\infty}\sum_{v = -\infty \setminus (0,0)}^{+\infty} ((ua+z_{x})^{2}+(va+z_{y})^{2}+h^{2} )^{-\beta},\]
\[ S_{v}=\sum_{u = -\infty}^{+\infty}\sum_{v = -\infty \setminus (0,0)}^{+\infty} ((ua+z_{x})^{2}+(va+z_{y})^{2}+h^{2} )^{-2\beta}. \]
From \cite{atc1}, for a given height to inter-LED separation ratio $h/a$, we can write a closed form expression for $S_{m}$ and $S_{v}$ as follows.
\begin{align}
S_{m}\approx S'_{m}= \frac{h^{2-2\beta}\pi}{a^{2}(\beta-1)} -\frac{1}{(z^{2}+ h^{2})^{\beta}} + \sum_{(w,f)\in\mathbb{A}}g_{m}(w,f),
\label{eqn:sm1}
\end{align}
\begin{align}
S_{v}\approx S'_{v}= \frac{h^{2-4\beta}\pi}{a^{2}(2\beta-1)} -\frac{1}{(z^{2}+ h^{2})^{2\beta}} + \sum_{(w,f)\in\mathbb{A}}g_{v}(w,f), 
\label{eqn:sv1}
\end{align}
where $g_{m}(w,f)=$
\begin{align*}
\frac{\mathbb{K}_{\beta-1}\big(\frac{2\pi h\sqrt{f^{2}+w^{2}}}{a}\big)\cos\big(\frac{2\pi wz_{x}}{a}\big)\cos\big(\frac{2\pi fz_{y}}{a}\big)}{\bigg(\frac{h}{2\pi\sqrt{f^{2}+w^{2}}}\bigg)^{\beta-1}2^{\beta-4} a^{\beta+1}\frac{\Gamma(\beta)}{\pi}},
\end{align*}
$g_{v}(w,f)=$
\begin{align*}
\frac{\mathbb{K}_{2\beta-1}\big(\frac{2\pi h\sqrt{f^{2}+w^{2}}}{a}\big)\cos\big(\frac{2\pi wz_{x}}{a}\big)\cos\big(\frac{2\pi fz_{y}}{a}\big)}{\bigg(\frac{h}{2\pi\sqrt{f^{2}+w^{2}}}\bigg)^{2\beta-1}2^{2\beta-4} a^{2\beta+1}\frac{\Gamma(2\beta)}{\pi}},
\end{align*}
and the set $\mathbb{A} \triangleq (\mathbb{Z}^{2}\cap([0,j]\times[0,l]))\setminus (0,0)$ over the set of integers $\mathbb{Z}^{2}$. For most of the practical use-cases, it is sufficient to choose only the terms corresponding to $j=l=1$, i.e., $(w=0,f=1), (w=1,f=0)$ and $(w=1,f=1)$. \par
Hence, the $P_{c}$ can now be extended from \eqref{eqn:pc1} as
\begin{align}
P_{c}&=\int_{-\frac{a}{2}}^{\frac{a}{2}}\int_{-\frac{a}{2}}^{\frac{a}{2}}\bigg[\int_{0}^{\eta}\frac{1/a^2}{\sqrt{2\pi}\sigma_{1}}\exp\bigg(\frac{-(x-\mu)^2}{2\sigma_{1}^{2}}\bigg)\d x\bigg] \d z_x \d z_y, \nonumber\\
&=\int_{-\frac{a}{2}}^{\frac{a}{2}}\int_{-\frac{a}{2}}^{\frac{a}{2}}\bigg(\frac{Erf\bigg(\frac{\eta-\mu}{\sqrt{2}\sigma_{1}} \bigg)+Erf\bigg(\frac{\mu}{\sqrt{2}\sigma_{1}}\bigg)}{2a^2}\bigg) \d z_x \d z_y, \nonumber\\
&\stackrel{(a)}{\approx}\int_{-\frac{a}{2}}^{\frac{a}{2}}\int_{-\frac{a}{2}}^{\frac{a}{2}}\frac{1}{2a^2}\bigg(Erf\bigg(\frac{\eta-pS'_{m}}{\sqrt{2}p(1-p)S'_{v}} \bigg)\nonumber\\
&\ \ \ \ \ \ \ \ \ \ \ \ \ \ \ \ \ \ \ \ \ \ \ \ \ \ +Erf\bigg(\frac{S'_{m}}{\sqrt{2}(1-p)S'_{v}}\bigg)\bigg) \d z_x \d z_y, 
\label{eqn:pc2}
\end{align}
where $Erf(x)=\int_{0}^{x}\frac{2e^{-t^2}}{\sqrt{\pi}}\d t$ is the standard Gaussian error function and $(a)$ follows from the approximation in \eqref{eqn:sm1} and \eqref{eqn:sv1}. 

\section{Simulation results}
The $P_c$ is validated using both numerical and analytical simulations at various values of $h/a\in\{3, 4, 5, 6\}$, and at different modulation probabilities $p\in\{0.3, 0.5, 0.8\}$ respectively in Fig.\ref{pc_03}, \ref{pc_05}  and \ref{pc_08}. Other optical parameters are shown in Table \ref{abc}. 

\begin{table}
\caption{Parameters considered for numerical simulations}
\label{abc}
%
\caption{At $p=0.5$, the variation of probability of coverage $P_{c}$ is plotted against the threshold $\theta$, for different values of the ratio $h/a$ of LED installation. The ratio $h/a$ is realized by assuming $a=0.5$m and $h\in\{1.5, 2.0, 2.5, 3.0\}$m.} 
\label{pc_05}
\end{figure} 

\begin{figure}[ht]
\centering
\definecolor{mycolor1}{rgb}{0.46600,0.67400,0.18800}%
\definecolor{mycolor2}{rgb}{0.30100,0.74500,0.93300}%
\definecolor{mycolor3}{rgb}{0.63500,0.07800,0.18400}%
\definecolor{mycolor4}{rgb}{0.00000,0.44700,0.74100}%
%
%
\caption{At $p=0.3$, the variation of probability of coverage $P_{c}$ is plotted against the threshold $\theta$, for different values of the ratio $h/a$ of LED installation. The ratio $h/a$ is realized by assuming $a=0.5$m and $h\in\{1.5, 2.0, 2.5, 3.0\}$m.} 
\label{pc_03}
\end{figure} 

\begin{figure}[ht]
\centering
\definecolor{mycolor1}{rgb}{0.46600,0.67400,0.18800}%
\definecolor{mycolor2}{rgb}{0.30100,0.74500,0.93300}%
\definecolor{mycolor3}{rgb}{0.63500,0.07800,0.18400}%
\definecolor{mycolor4}{rgb}{0.00000,0.44700,0.74100}%
%
%
\caption{At $p=0.8$, the variation of probability of coverage $P_{c}$ is plotted against the threshold $\theta$, for different values of the ratio $h/a$ of LED installation. The ratio $h/a$ is realized by assuming $a=0.5$m and $h\in\{1.5, 2.0, 2.5, 3.0\}$m.} 
\label{pc_08}
\end{figure} 

In Fig.\ref{pc_05} for $p=0.5$, the analytical simulations for the probability of coverage show that the $P_c$ over the central attocell drops steadily with increase in the SINR threshold $\theta$. But when the ratio $h/a$ increases, this drop in $P_c$ occurs at higher values of $\theta$. For instance at $h/a=3$, $P_c=0.6$ is achieved when $\theta=-6.55$dB; and when $h/a$ is increased to $4$, $P_c=1$ which later drops with an increase in $\theta$ beyond $-6$dB. This also means that for a fixed inter-LED spacing $a$, more is the height $h$ of LED installation, a better coverage probability occurs even for lower values of the SINR at the PD. Parallelly, these analytical results have been validated with appropriate numerical simulations which are tight and are drawn neither with any Gaussian approximation nor the series approximation from \cite{atc1} for $S_{m}$ and $S_{v}$.\par
A similar trend can be observed in Fig.\ref{pc_03} for $p=0.3$, and in Fig. \ref{pc_08} for $p=0.8$ where $P_c$ drops steadily with increase in $\theta$. All the same, this range of $\theta$ shifts to a lower threshold band as the modulation probability $p$ increases. This is true because, as $p$ the probability that an LED will be a LiFi source increases, the attocell network will have more LiFi sources. This implies more co-channel interference and as a result lesser probability of coverage $P_c$. 

\section{Conclusion}
This work derived an exact expression for probability of coverage in a LiFi attocell network that is essentially thinned out of a deterministic one. Validations with numerical simulations are provided. Analysis of joint transmission schemes shall form a part of the future work.

\bibliographystyle{IEEEtran}
\bibliography{IEEEfull,referencecov.bib}

\begin{thebibliography}{1}
\providecommand{\url}[1]{#1}
\csname url@samestyle\endcsname
\providecommand{\newblock}{\relax}
\providecommand{\bibinfo}[2]{#2}
\providecommand{\BIBentrySTDinterwordspacing}{\spaceskip=0pt\relax}
\providecommand{\BIBentryALTinterwordstretchfactor}{4}
\providecommand{\BIBentryALTinterwordspacing}{\spaceskip=\fontdimen2\font plus
\BIBentryALTinterwordstretchfactor\fontdimen3\font minus
  \fontdimen4\font\relax}
\providecommand{\BIBforeignlanguage}[2]{{%
\expandafter\ifx\csname l@#1\endcsname\relax
\typeout{** WARNING: IEEEtran.bst: No hyphenation pattern has been}%
\typeout{** loaded for the language `#1'. Using the pattern for}%
\typeout{** the default language instead.}%
\else
\language=\csname l@#1\endcsname
\fi
#2}}
\providecommand{\BIBdecl}{\relax}
\BIBdecl

\bibitem{atc1}
A.~Surampudi and R.~K. Ganti, ``{Interference Characterization in Downlink
  Li-Fi Optical Attocell Networks},'' \emph{Journal of Lightwave Technology},
  vol.~36, no.~16, pp. 3211--3228, 2018.

\bibitem{hass}
{Haas, Harald}, ``{LiFi: Conceptions, misconceptions and opportunities},'' in
  \emph{{Photonics Conference (IPC), 2016 IEEE}}.\hskip 1em plus 0.5em minus
  0.4em\relax IEEE, 2016, pp. 680--681.

\bibitem{cov1}
L.~Yin and H.~Haas, ``{Coverage Analysis of Multiuser Visible Light
  Communication Networks},'' \emph{IEEE Transactions on Wireless
  Communications}, vol.~17, no.~3, pp. 1630--1643, 2018.

\bibitem{lit1}
C.~Chen, D.~A. Basnayaka, and H.~Haas, ``{Downlink Performance of Optical
  Attocell Networks},'' \emph{Journal of Lightwave Technology}, vol.~34, no.~1,
  pp. 137--156, 2016.

\bibitem{lit2}
L.~Yin and H.~Haas, ``{A Tractable Approach to Joint Transmission in Multiuser
  Visible Light Communication Networks},'' \emph{IEEE Transactions on Mobile
  Computing}, p. Early access.

\bibitem{lit3}
C.~Chen and H.~Haas, ``{Performance Evaluation of Downlink Cooperative
  Multipoint Joint Transmission in LiFi Systems},'' in \emph{Globecom Workshops
  (GC Wkshps), 2017 IEEE}.\hskip 1em plus 0.5em minus 0.4em\relax IEEE, 2017,
  pp. 1--6.

\bibitem{nlos1}
T.~Komine and M.~Nakagawa, ``{Fundamental Analysis for Visible-Light
  Communication System Using LED Lights},'' \emph{IEEE Transactions on Consumer
  Electronics}, vol.~50, no.~1, pp. 100--107, 2004.

\bibitem{nlos2}
J.~Grubor, S.~Randel, K.-D. Langer, and J.~W. Walewski, ``{Broadband
  Information Broadcasting Using LED-Based Interior Lighting},'' \emph{Journal
  of Lightwave Technology}, vol.~26, no.~24, pp. 3883--3892, 2008.

\bibitem{nlos3}
L.~Zeng, D.~C. O'Brien, H.~Le~Minh, G.~E. Faulkner, K.~Lee, D.~Jung, Y.~Oh, and
  E.~T. Won, ``{High Data Rate Multiple Input Multiple Output (MIMO) Optical
  Wireless Communications Using White LED Lighting},'' \emph{IEEE Journal on
  Selected Areas in Communications}, vol.~27, no.~9, 2009.

\end{thebibliography}

\end{document}